\documentclass[]{spie}  

\usepackage{amsmath,amsfonts,amssymb}
\usepackage{graphicx}
\usepackage{subcaption}
\usepackage{float}
\usepackage[table, dvipsnames,svgnames,x11names]{xcolor}
\usepackage[colorlinks=true, allcolors=blue]{hyperref}

\usepackage{subcaption}
\usepackage{tikz}

\newcommand*\arcsec{$^{\prime\prime}$}

\title{Technical requirements flow-down for the concept design of the novel 50-meter Atacama Large Aperture Submm Telescope (AtLAST)}

\author[a]{Matthias Reichert}
\author[a]{Martin Timpe}
\author[b]{Hans Kaercher}
\author[c]{Tony Mroczkowski}
\author[a]{Manuel Groh}
\author[a]{Aleksej Kiselev}
\author[d]{Claudia Cicone}
\author[e]{Patricio Gallardo}
\author[f]{Roberto Puddu}
\author[g]{Pamela Klaassen}

\affil[a]{OHB Digital Connect, Weberstra\ss e 21, D-55130 Mainz, Germany}
\affil[b]{Independent Consultant, Kirchgasse 4, D-61184 Karben, Germany}
\affil[c]{European Southern Observatory, Karl-Schwarzschild-Str.\ 2, Garching 85748, Germany}
\affil[d]{Institute of Theoretical Astrophysics, University of Oslo, P.O. Box 1029, Blindern, 0315 Oslo, Norway}
\affil[e]{Kavli Institute for Cosmological Physics, University of Chicago, Chicago, IL, 60637, USA}
\affil[f]{Instituto de Astrofísica and Centro de Astro-Ingeniería, Facultad de Física, Pontificia Universidad Católica de Chile, Santiago, Chile}
\affil[g]{UK Astronomy Technology Centre, Royal Observatory Edinburgh, Blackford Hill, Edinburgh EH9 3HJ, UK}

\authorinfo{Further author information: (Send correspondence to M.R.)\\
M.R.: E-mail: matthias.reichert@ohb.de}

\begin{document} 
\maketitle

\begin{abstract}
The Atacama Large Aperture Submm Telescope (AtLAST) is a concept for a novel 50-meter class single-dish telescope operating at sub-millimeter and millimeter wavelengths (30-950 GHz). The telescope will provide an unprecedentedly wide field of view (FoV) of 1-2 degree diameter with a large receiver cabin housing six major instruments in Nasmyth and Cassegrain positions. The high observing frequencies, combined with the scanning operation movements with up to 3 deg/second, place high demands on the accuracy and stability of the optical and structural components. The design features the introduction of a rocking chair type mount with an iso-statically decoupled main reflector backup structure and an active main reflector surface with a high precision metrology system. The planned site location is in the Chilean Atacama Desert at approximately 5050 meters above sea level, near Llano de Chajnantor. This paper gives an overview of the optical, structural, and mechanical design concepts. It explains the flow-down from key science requirements to technical design decisions as well as showing design analogies from other existing large radio, (sub-)mm, and optical telescopes.

\end{abstract}


\section{INTRODUCTION}
\label{sec:intro}  

The Atacama Large Aperture Submm Telescope (AtLAST)\footnote{\href{https://atlast-telescope.org/}{https://atlast-telescope.org/}} is a project undergoing a thorough conceptual design study, funded by EU Horizon 2020 program, running from March 2021 through August 2024. The study consortium consists of the University of Oslo (UiO, Coordinator), the European Southern Observatory (ESO), UK Research and Innovation (UKRI), University of Hertfordshire (UNIHERT) and OHB Digital Connect GmbH (OHB DC) as an industrial partner. The study's goals are to set the path for the future observatory by defining the science cases, developing a telescope design, finding a suitable site, generating first operation concepts, investigating the sustainable energy supply, and assessing potential governance models for the future realization of the project. 

Results describing the actual telescope design have been presented to the astronomical community before by Mroczkowski et al.\ 2023, 2024 \cite{Mroczkowski2023, Mroczkowski2024}, while Viole et al.\ 2023 \cite{Viole2023} and Kiselev et al.\ 2024 \cite{Kiselev2024} describe the efforts to make AtLAST one of the first facilities to consider sustainable power and energy saving concepts from its inception. 
The present work focuses on describing the requirements flow down from science cases to engineering budgets and presents a selection of technical solutions derived from the design iterations.

\section{Science enabled by AtLAST}\label{sec:Science}

Electromagnetic emission from the Galactic and extragalactic sky peaks at sub-millimeter and millimeter (hereafter, (sub-)mm) wavelengths, containing a wealth of information about the Universe inaccessible to other bands.
For instance, (sub)-mm astronomical observations allow astronomers to probe matter that is largely invisible at other wavelengths due it being extremely cold (temperature $T<10^3 \, \rm K$), too warm or hot ($T \geq 10^5 \, \rm K$), or -- for objects at large cosmological distances -- having its bright spectral lines redshifted out of the ultraviolet, optical, and infrared bands. 

Yet our view of the (sub)-mm sky has largely been limited to low, arcminute resolutions or to small areas (tens of square arcminutes).
AtLAST's planned (sub-)mm observational capabilities are uniquely suited to change this, providing a powerful window into a broad range of astrophysics spanning all states of matter. 
The high surface brightness sensitivity, angular resolution, mapping speed, imaging dynamic range, wavelength coverage, and solar capabilities of the AtLAST concept result directly from the needs of delivering the following science cases, none of which can be achieved through current or funded future facilities:
\begin{enumerate}

    \item To perform the deepest, widest (100-1000 deg$^2$), and most complete imaging and spectroscopic surveys of the Galactic and extragalactic sky at a few arcsecond resolution ($\approx 3$\arcsec\ at 500~GHz), beating the confusion limits of earlier experiments\cite{Klaassen2020, Ramasawmy2022, vanKampen2024atlast}. The goal for the Galactic observations is to understand and fully characterize this complex ecosystem; that is, the interplay between gravity, turbulence, and magnetic fields at scales from protostellar cores up to giant molecular cloud structures, and to detect the faint planetary belts and Kuiper belt analogues forming around stars\cite{Klaassen2024atlast}. These findings will directly inform extragalactic studies with AtLAST, which aim to fully characterize the feedback cycles and evolution of galaxies near and far. For this, AtLAST aims to achieve the most statistically complete, homogeneous, and unbiased census of the star forming galaxy population from the nearby to the distant Universe,\cite{Liu2024atlast, vanKampen2024atlast} resolving the cosmic infrared background, and producing a dataset with an immense legacy value, comparable to that of optical/near-IR surveys such as the Sloan Digital Sky Survey, the Dark Energy Survey, or the Vera Rubin Observatory's Legacy Survey of Space and Time. 

    \item To detect and image the diffuse and low surface brightness signals of the gas in the dark matter haloes of galaxies, galaxy groups, and galaxy clusters. We aim to constrain the morphology, kinematics, and chemistry of cold gaseous structures within the large-scale interstellar medium (ISM)\cite{Liu2024atlast} and circumgalactic medium (CGM)\cite{Lee2024atlast, Schimek2024} of galaxies through molecular and atomic line emissions. In parallel, we aim to perform the most sensitive sub-arcmin resolution observations of the Sunyaev-Zel’dovich (SZ) effect, deriving the thermodynamics and kinematics of the warm/hot ionized gas phase in massive cosmic structures \cite{DiMascolo2024}. These observations can provide a complete view of the elusive cycling of baryons in and out of galaxies, which shapes and controls their evolution. 

    \item To study the thermal structure and heating of the highly dynamic solar chromosphere, including the variability of flares on time scales below one minute, and so unlock the potential diagnostic power of \mbox{(sub-)mm} observations to understand solar/stellar activity, its impact on (exo-)planetary systems, and to reveal a new transient population \cite{Cordiner2024atlast,orlowskischerer2024atlast,Wedemeyer2024atlast}. AtLAST's mm/submm view of the time variable sky will complement and complete the view given by low frequency radio, optical/near-IR, neutrino, high-energy (Cherenkov radiation), and gravitational wave probes.
\end{enumerate}

For more detail on the science requirements and motivations for AtLAST, see Booth et al.\ 2024\cite{Booth2024} in this volume, the AtLAST key science driver summary report by Booth et al.\ 2024b\cite{Booth2024b}, the summary of use cases presented previously in Ramasawmy et al.\ 2022\cite{Ramasawmy2022}, the case for mm-wave very long baseline interferometry (mm-VLBI)\cite{Akiyama2023} and the recent suite of eight science cases recently submitted for publication\cite{Cordiner2024atlast, DiMascolo2024, Klaassen2024atlast, Lee2024atlast, Liu2024atlast, orlowskischerer2024atlast, vanKampen2024atlast, Wedemeyer2024atlast}.\footnote{The AtLAST science cases can be found in the Open Research Europe collection, here: \url{https://open-research-europe.ec.europa.eu/collections/atlast}.}

\section{Requirements Flow Down from the Science Goals}
AtLAST’s science goals result in a number of fundamental requirements for the telescope design, summarized in Table~\ref{tab:requirements}. To pursue groundbreaking new science, AtLAST needs to have a 50~m diameter dish with high throughput and a FoV subtending at least one degree. These characteristics will enable astronomers to map the sky $>10^5 \times$ faster than ALMA could ever achieve, down to low noise levels and high resolutions unprecedented for any other single-dish facility, current or planned. The design of AtLAST will need to accommodate more than five future generation of advanced (sub-)mm instruments, including highly multiplexed high-resolution spectrometers, continuum cameras, and ultra-wideband integral field units. These ambitious requirements make the development and construction of AtLAST a significant, but not insurmountable, technical challenge.

\begin{table}
    \centering
\caption{Summary of the key technical requirements for AtLAST, adapted from Mroczkowski et al.\ 2024.\cite{Mroczkowski2024}}
\label{tab:requirements}
    \begin{tabular}{|l|c|}
      \hline             
      \textbf{Parameter} & \textbf{Value} \\    
\hline                       
Wavelength ($\lambda$) range		    & 0.3-10 mm\\ \hline 
Primary Mirror Diameter                 & 50 m\\ \hline 
Field of View (FoV)		                & 2$^\circ$ (1$^\circ$)
\\ \hline 		
Number of Instruments                   & $\geq 5$             \\ \hline 		
Effective Focal Length		            & $\approx 100$  m    \\ \hline 
Total Collecting Area		            & $\approx 2000$  m$^2$\\ \hline 
Optical Half Wavefront Error    	    & 20-25~$\mu$m         \\ \hline 
Surface Coating for Solar observations		                & similar to ALMA      \\ \hline 
Optical Design                          & Cassegrain-Nasmyth   \\ \hline 
Blind Pointing accuracy                 & 2.5\arcsec         \\ \hline 
Scan Speed                              & 3$^\circ \rm s^{-1}$ \\ \hline 
Acceleration                            & 1$^\circ \rm s^{-2}$ \\ \hline 
Elevation (EL) range                    & $20 ^\circ - 90^\circ$ \\ \hline 
Azimuthal (AZ) range                    & $\pm 270^\circ$ \\ \hline 
Mount Type                              & AZ-EL       \\	\hline 
\end{tabular}
\end{table}

From the telescope designer point of view these requirements result in three main technical challenges, compared with previous telescopes of similar size and scope:
\begin{enumerate}
    \item 2 deg field of view (FoV) and multi-instrument use: The FoV and imaging requirement leads to a spatial dynamic range comparable to optical telescopes.  By way of comparison, most existing (sub-)mm wavelength telescopes observing with only one to a handful of spatial elements, with a few exceptional experiments now reaching 10's of thousands of imaging detectors; AtLAST will ultimately be able to accommodate of order several million ($\mathcal{O}(>10^6)$) imaging detectors, or beams, while multi-chroic or spectrometers would dramatically increase this.
    Further, AtLAST is designed to be a long-lived facility instrument, serving many science cases, foreseen and unforeseen, from across a broad community.  
    The result is that AtLAST must be able to host multiple extremely massive (8-30 ton) astronomical instruments,\cite{AtLAST_memo_3} which drives the need for a large and accessible receiver bay. Switching rapidly between these instruments shall be possible during telescope operation, ensuring flexible and dynamic operations (e.g.\ as weather conditions change). 
    \item 20~$\mu$m rms half wavefront error (HWFE): The observation wavelengths and the field of view for which AtLAST is designed lead to a half wavefront error requirement of 20 $\mu$m, which is very demanding for such a large (sub-)mm telescope. Beyond a sophisticated backup structure based on homology principles as described in Kärcher \& Baars 2000,\cite{KaercherBaars2000} this can only be achieved by active reflector optics that is capable to react to slow transient changes of environmental disturbances. 
    \item Blind pointing error of 2.5\arcsec: The requested uncorrected pointing error budget is very demanding for a large telescope primary and secondary mirror structure that is directly exposed to the environment and not protected by a dome or enclosure. It is expected that this can be accomplished by a sophisticated combination of measurement systems. Those do mainly consist of a stiff elevation structure, a system to measure and compensate changes in the telescope axes orientation and actively aligned optical elements.
\end{enumerate}
The following sections describe how the requirements flow down into design decisions, taking into account the experience from existing telescope facilities.

\section{Large field of view and multi-instrument use}

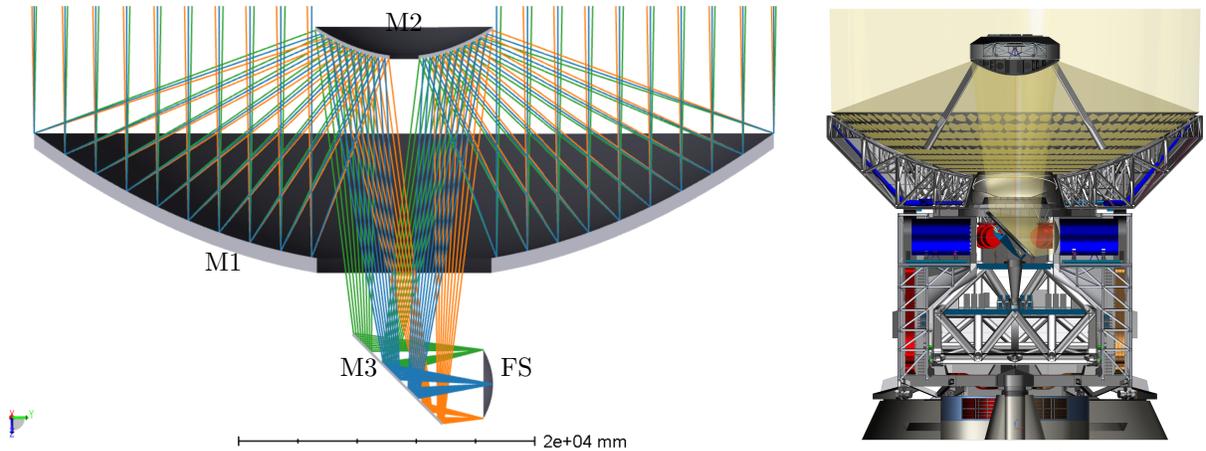
\begin{figure}[tbh]
        \centering
    \begin{subfigure}[b]{0.64\hsize}
    		\centering
            \begin{tikzpicture}
    \draw (0, 0) node[inner sep=0] {\includegraphics[width=\hsize, trim={0.15cm 0.15cm 0.15cm 1.5cm},clip]{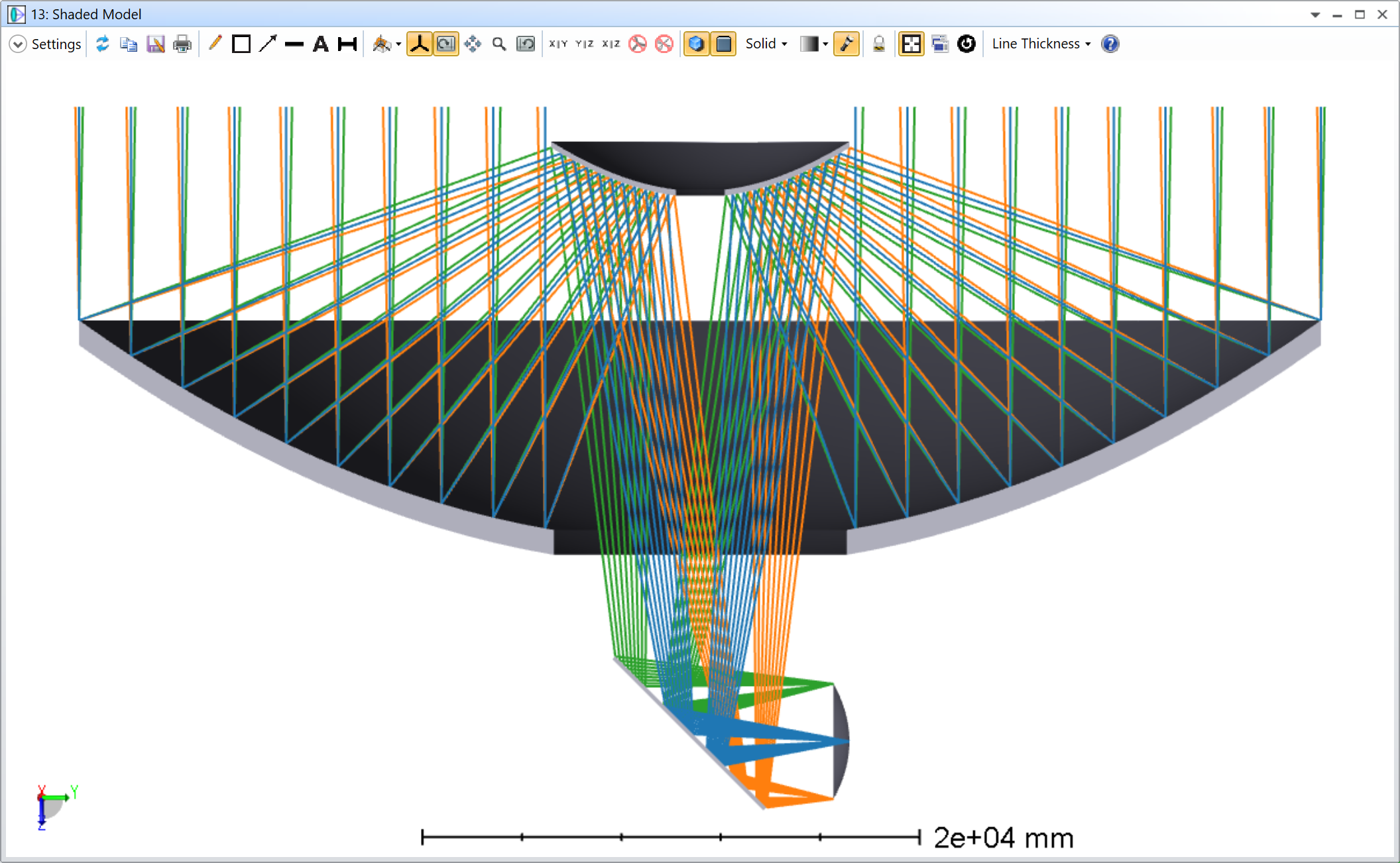}};
    \draw (-2.4,  -0.6) node {M1};
    \draw (0,  2.6) node {M2};
    \draw (-0.6, -2.0) node {M3};
    \draw (1.5, -2.0) node {FS};
\end{tikzpicture}
    \end{subfigure}  
    \includegraphics[height=5.9cm, trim={28cm 2cm 42cm 3cm},clip]{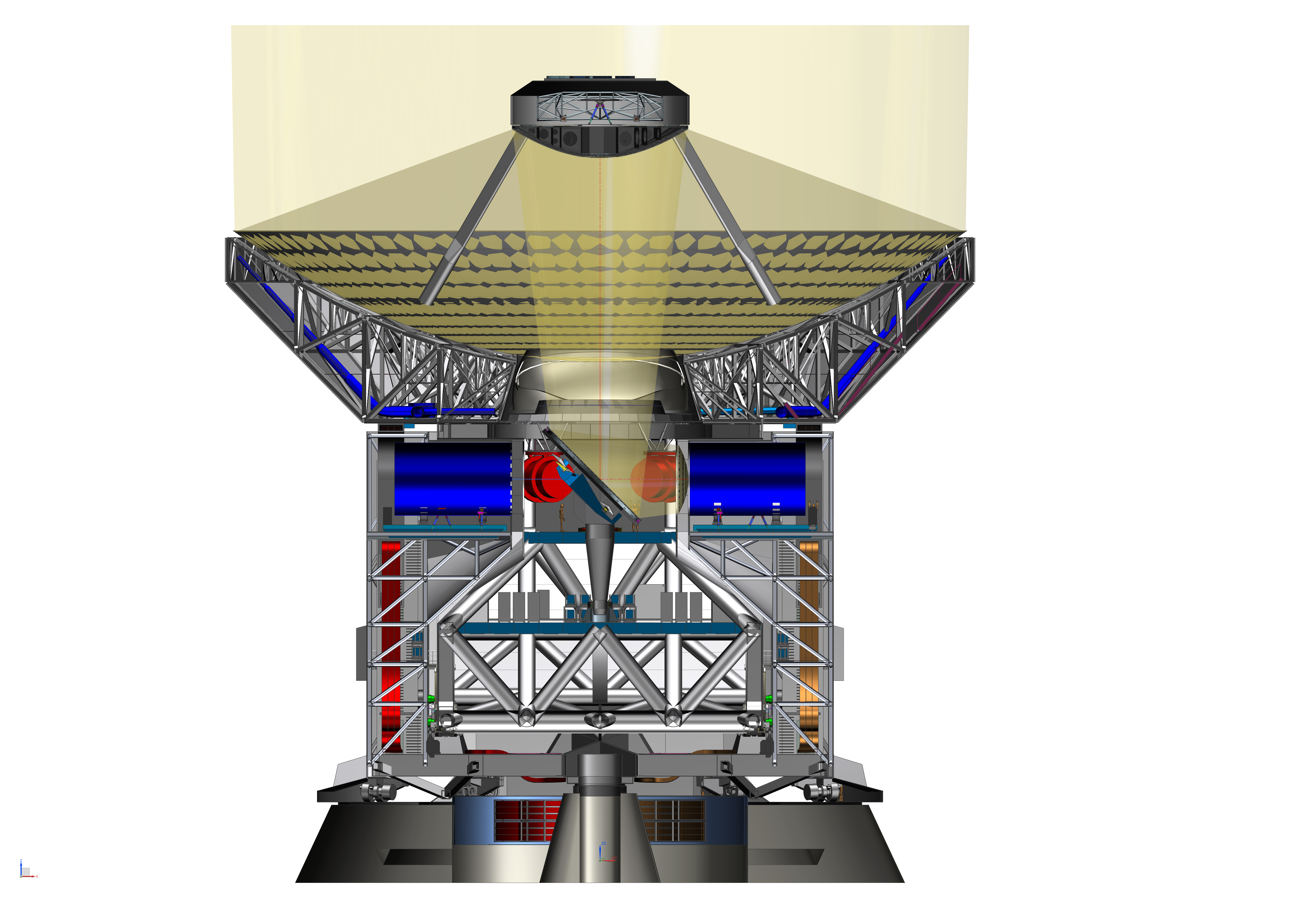}
        \caption{Left: The optical layout for AtLAST, reproduced from Mroczkowski et al.\ 2024\cite{Mroczkowski2024}. The primary, secondary, and tertiary mirrors (M1, M2 and M3, respectively) concentrate and direct light from the sky onto the focal surface (labeled ``FS'').
        Right: A rendering of the AtLAST CAD model including the optical path in gold.  The CAD model, shown in Figures~\ref{fig:CAD Overview} \& \ref{fig:exploding_floor}, is described in more detail in this section as well as Mroczkowski et al.\ 2024\cite{Mroczkowski2024}.
        }
        \label{fig:atlast_optics}
\end{figure}

The optimal optical concept for AtLAST was developed by R.\ Hills in the first AtLAST memo.\cite{AtLAST_memo_1} In this memo, a baseline or ``preferred'' design is identified as likely the most feasible one that can achieve the goals of AtLAST. This is a Ritchey-Chrétien arrangement based on two imaging mirrors (M1 and M2). The M3 Nasmyth mirror is flat and only folding the optical beam into the direction of the instrumentation. Ideally, M3 makes no contribution to the imaging features of the overall optical system.
Fig.~\ref{fig:atlast_optics} shows the optical design, reproduced from Mroczkowski et al.\ 2024\cite{Mroczkowski2024}. For more details about the optical design, see Gallardo et al.\ 2024\cite{Gallardo2024} and Puddu et al.\ 2024\cite{Puddu2024} (submitted, these proceedings). We note that the focal surface reaches diameters of up to 4.7~m, which will result in larger instruments, either monolithic or as arrays of smaller instruments, than those used in the current state of the art (e.g., the camera array concept developed in Gallardo et al.\ 2024\cite{Gallardo2024}).

\begin{figure}[tbh]
    \centering
    \includegraphics[trim={1.5cm 0.3cm 1.5cm 1.0cm},clip, width=0.8\linewidth]{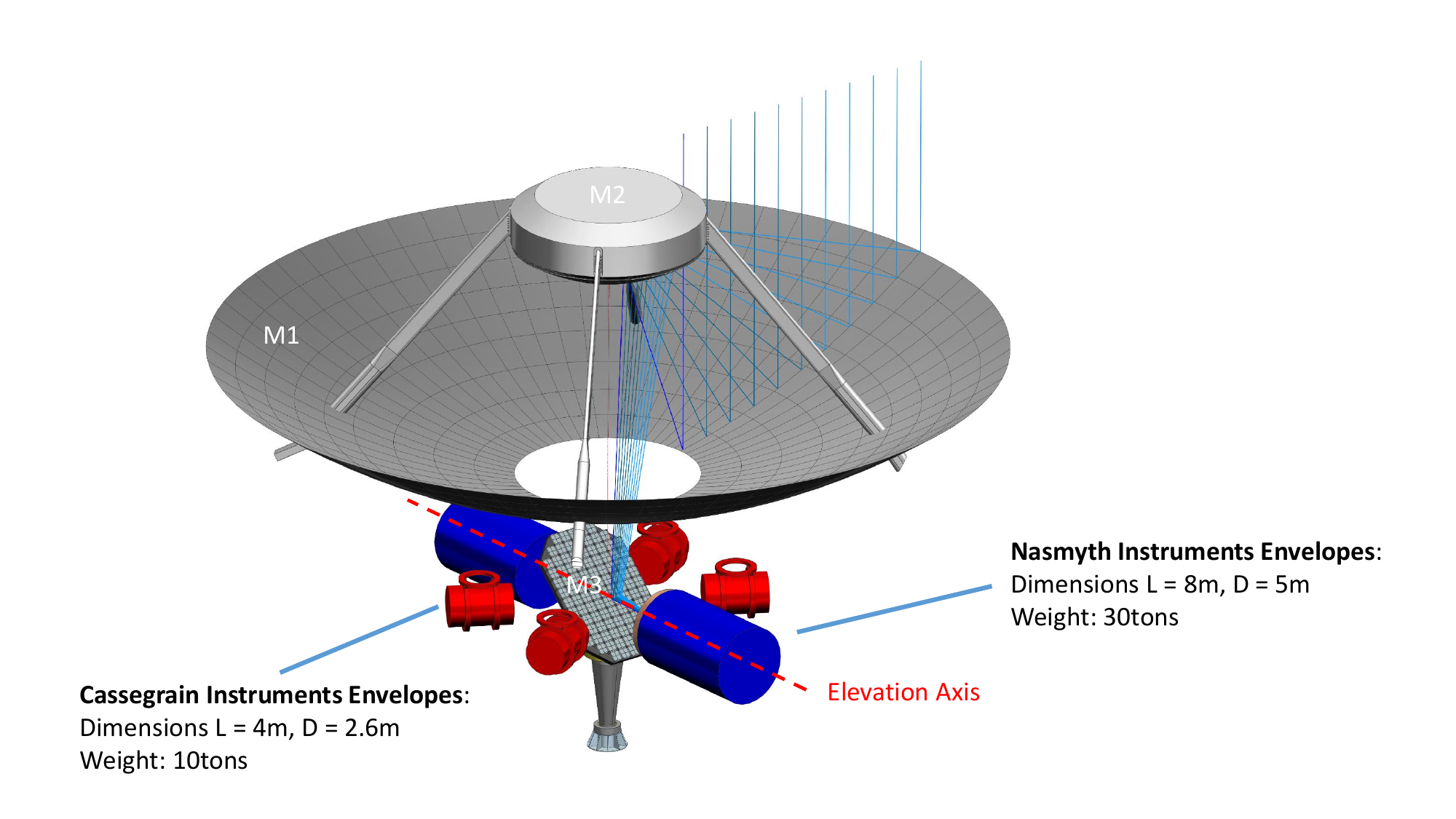}
    \caption{Optical arrangement of reflectors and astronomical instruments arranged around the folding tertiary mirror (M3). The larger, Nasmyth instruments, which are located along the elevation axis, are shown in blue, and Cassegrain instruments, which co-rotate with the elevation wheel, are shown in red. As in Fig.~\ref{fig:atlast_optics}, an example ray trace is also shown.}
    \label{fig:instrument arrangement}
\end{figure}

A strong demand from AtLAST’s scientific goals is the ability to carry several instruments of different kinds and sizes as well as switching between them in a short period of time. Also, at least two of the instruments shall be Nasmyth instruments that do not rotate with the telescope structure in elevation, thus being located in the elevation axis of the telescope. Fig.~\ref{fig:instrument arrangement} shows the resulting arrangement of the instrument bays around a rotatable M3 to switch between them. The instrument diameters are up to 5~m for the Nasmyth instruments, and up to 3~m diameter for the Cassegrain instruments.

\begin{figure}[tbh]
    \centering
    \includegraphics[trim={2.5cm 1.9cm 3.0cm 2.2cm},clip, width=0.8\linewidth]{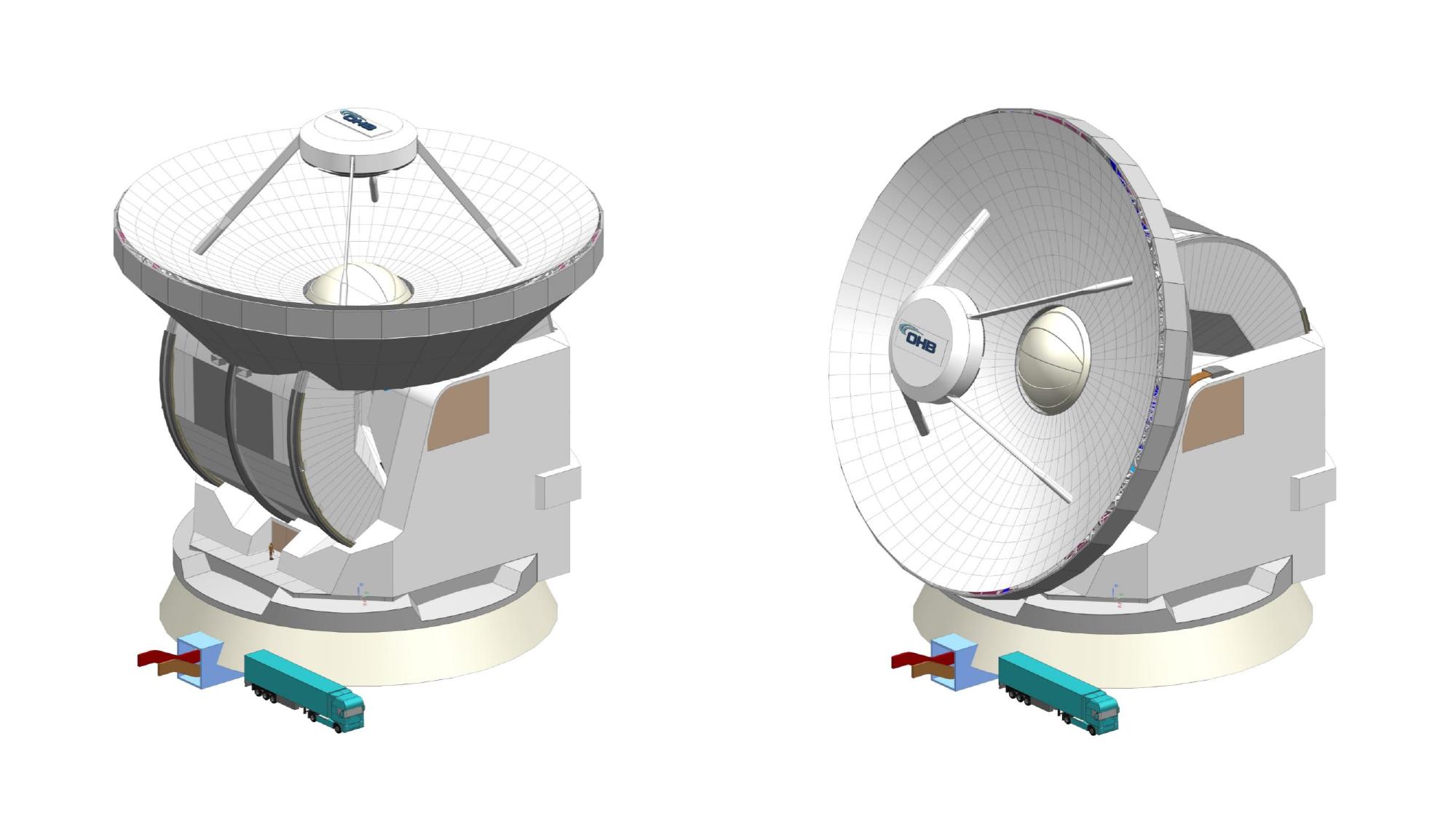}
    \caption{AtLAST external view in 90$^\circ$ position (left) and 20$^\circ$ position (right), with a freight truck shown for scale. Fig.~\ref{fig:exploding_floor} reveals the internal substructures. }
    \label{fig:CAD Overview}
\end{figure}

\begin{figure}[tbh]
    \begin{subfigure}[t]{0.66\hsize}
    		\centering
            \includegraphics[trim={0.8cm 0.8cm 0.8cm 1.2cm}, clip, width=\linewidth]{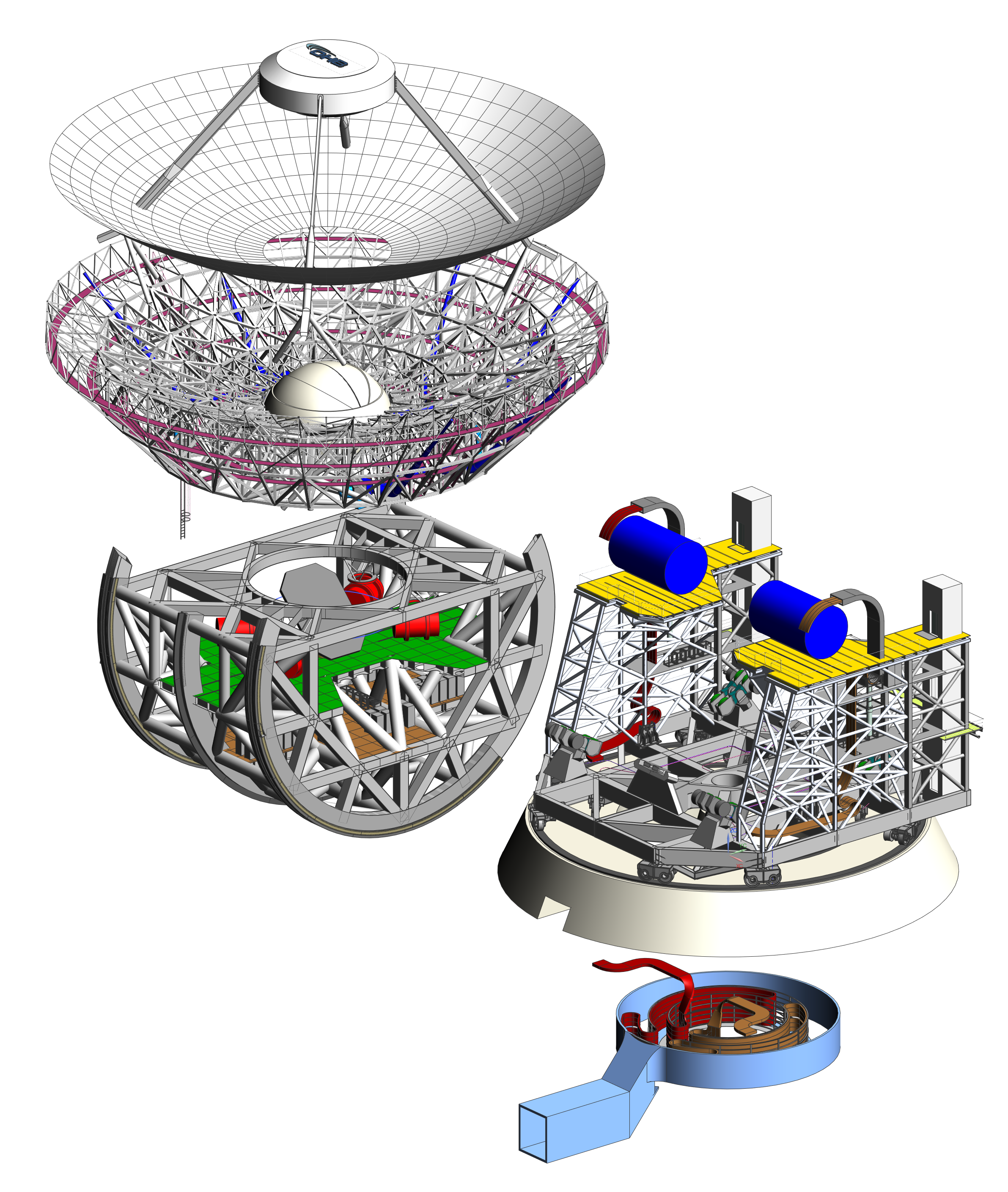}
    \end{subfigure} 
    \begin{subfigure}[b]{0.34\hsize}
            \includegraphics[trim={5.4cm -14.0cm 7.0cm 0.0cm}, clip, width=\linewidth]{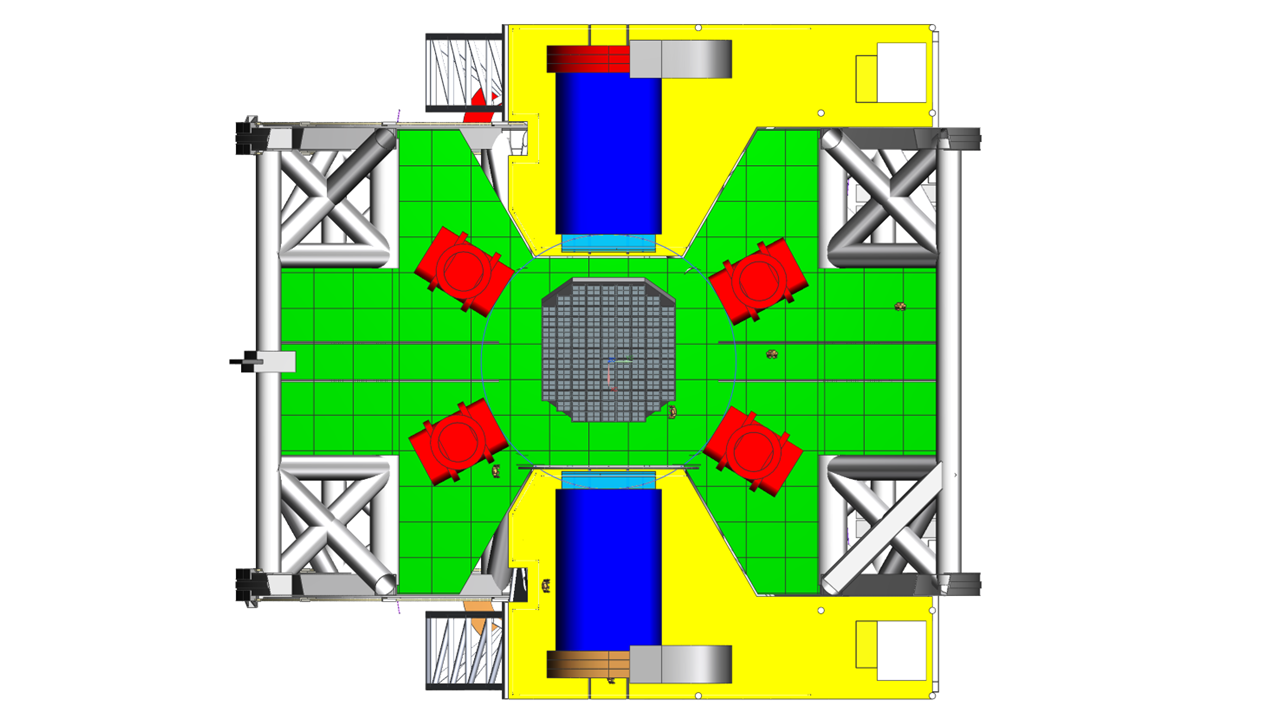}
    \end{subfigure} 
        \caption{
        Left: Explosion view with reflectors and elevation wheel, with the azimuthal track and cable wrap below it offset to the lower right.
        Right: Top view of the instrument platforms, with colors as denoted in the left panel.  As in Fig.~\ref{fig:instrument arrangement}, this image shows the 
        Nasmyth instruments in blue and Cassegrain instruments in red, arranged around the folding tertiary mirror (dark gray).
        }
        \label{fig:exploding_floor}
\end{figure}

This instrument arrangement concept evokes current designs for large optical telescopes like the Giant Magellan Telescope (GMT) and the Extremely Large Telescope (ELT)\cite{Tamai2014ELT}. Thus, it is obvious to follow similar structural design approaches. The final outcome after several design iterations is the rocking chair design as shown in Figures~\ref{fig:CAD Overview} \& \ref{fig:exploding_floor}. The 50~m reflector rests on a so-called elevation wheel providing a stiff support structure and carrying in its inner large instrument room the Cassegrain instruments as well as a rotating M3. The two large Nasmyth instruments are located on two fixed towers mounted on the azimuth structure, one on each side of the elevation wheel. The elevation wheel runs on a bogie and track configuration on the azimuth structure. The azimuth structure rotates around a central bearing, while the vertical loads are carried by twelve azimuth bogies running on two concentric tracks.
All structural elements are covered by a thermal cladding for temperature and ventilation management. 
More details about the subsystem design can be found in Mroczkowski et al.\ 2024 \cite{Mroczkowski2024}.

\section{20 microns RMS Half Wavefront Error}

\subsection{Half Wavefront Error Budget}

The key telescope requirement is to achieve a 20~$\mu$m rms half wavefront error (HWFE) error, derived from the observation wavelengths of 350~$\mu$m.  
This places high demands on the surface accuracy of all reflecting surfaces and the optical quality, determined by the HWFE, as well as the pointing. The requested accuracy of the reflectors is about four times higher than the one specified and achieved for the Large Millimeter Telescope (LMT) and in the range of the accuracy requested for the 12m ALMA antennas. The primary reflector HWFE being one of the main contributors to the full error budget, the proposed design for the M1 surface is therefore based on proven ideas and also experiences with the two aforementioned predecessors. 

It should be immediately obvious that active compensation techniques are needed in order to reach the demanded optical surface accuracy and pointing precision. A constantly operational, active main reflector surface and a sophisticated metrology system are inevitable to compensate various disturbances which affect the optical quality and the pointing during the telescope operation.

\begin{figure}
    \centering
    \includegraphics[trim={1.5cm 8.5cm 2.0cm 2.0cm}, clip, width=1\linewidth]{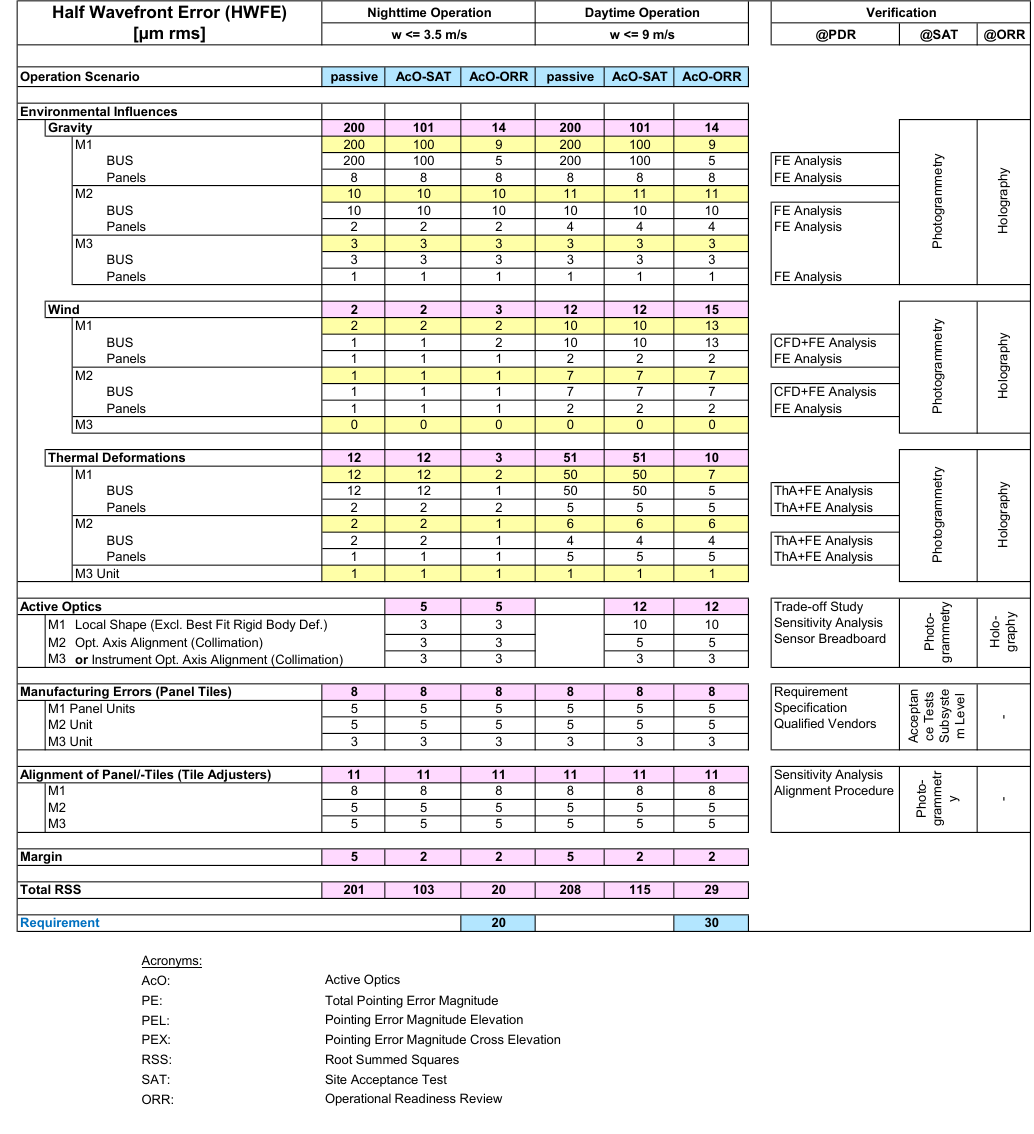}
    \caption{Half wavefront error (HWFE) budget}
    \label{fig:HWFE Budget}
\end{figure}

Fig.~\ref{fig:HWFE Budget} shows the flow down of a maximal 20 $\mu$m  HWFE requirement to the subsystem level. The table is established in a top-down manner, which means that individual contributions of the subsystems are distributed in a way that the overall outcome just fulfills the HWFE requirement. The table separates into two operation modes: 
\begin{enumerate}
    \item Nighttime operations: Short wavelength observations ($\lesssim 600 \mu$m) due to preferred environmental conditions (low humidity, lower winds, reduction of thermal structure deformation due to sunlight exposure) that require the low half wavefront error of 20~$\mu$m rms (or better). The highest accuracy in operation is ensured to wind speeds of 3.5 m/s. However, operation above this speed is possible for observation in lower frequency ranges.
    \item Daytime operations: Long wavelength observations that can tolerate a slightly degraded half wavefront error at higher wind speed of up to 9 m/s. It is assumed that it may be reduced to 30~$\mu$m rms. 
\end{enumerate}

The error contributors are separated into
\begin{itemize}
    \item environmental influences from gravity, wind, and thermal conditions,
    \item errors introduced by the manufacturing processes of the related subsystems,
    \item alignment errors introduced by the alignment procedures during assembly on site, and
    \item active optics residual alignment errors.
\end{itemize}

The verification methods adopted during detailed design and during the final acceptance are indicated in two columns on the right side of the table. We assume consecutive steps of parameter verification: in the design phase of the telescope most requirements will be verified by simulation. The ultimate verification happens after construction of the telescope. 
At the site acceptance testing (SAT) the telescope contractor will demonstrate the achievement with specified methods like photogrammetry and holography, whose accuracy and precision limits will define the limits for the verification. 
Here ends the telescope contractor’s responsibility and the telescope is handed over to the operators.
In most cases that performance verification does not correspond to the final performance of the telescope. 
The telescope will reach its maximum performance after a calibration phase by the observatory operator and scientists, applying the astronomical models on the tracking algorithms and using the actual scientific instruments. This calibration phase ends with an Operational Readiness Review (ORR), realizing the full operational performance.

From analysis and considerations of the structural deformations, it becomes obvious that the ambitious goal of 20 $\mu$m rms for short wavelengths requires the use of active correction systems: 
\begin{enumerate}
    \item Active optics (AcO): What we define as AcO is a system that actively adjusts the panel orientation to maintain the HWFE (compensation of the flexibility of the main reflector) and measures and controls the alignment of the optical components (reflecting surfaces and instrument as quasi rigid bodies) in the presence of gravity and environmental loads in a closed loop with a bandwidth between 0.1-0.3~Hz (i.e.\ active corrections on timescales $\approx 3-10$~seconds).
    \item Flexible Body Compensation (FBC): A system that directly or indirectly measures the orientation of the elevation and azimuth axes and applies pointing corrections only by means of commanding the main axes drives (e.g. measurement and compensation of deformations of the elevation wheel and the azimuth rotating structure).  We note any components within the optical assembly (i.e.\ the reflector panel segments) are controlled by the AcO described above.
\end{enumerate}
The HWFE table columns show a separation between the passive, structural deformation behavior without corrections by the AcO on the one hand and the corrections by the AcO to achieve the final accuracy on the other hand. 

\subsection{Passive Measures}
The active system is inevitable; nevertheless it is a reasonable and proven approach to make the passive system as good as possible before implementing active correction. Thus, the passive structural elements must provide a good base without large scale use of extremely expensive material (e.g. CFRP backup structure).

\begin{figure}[htb]
    \centering
    \includegraphics[width=1\linewidth]{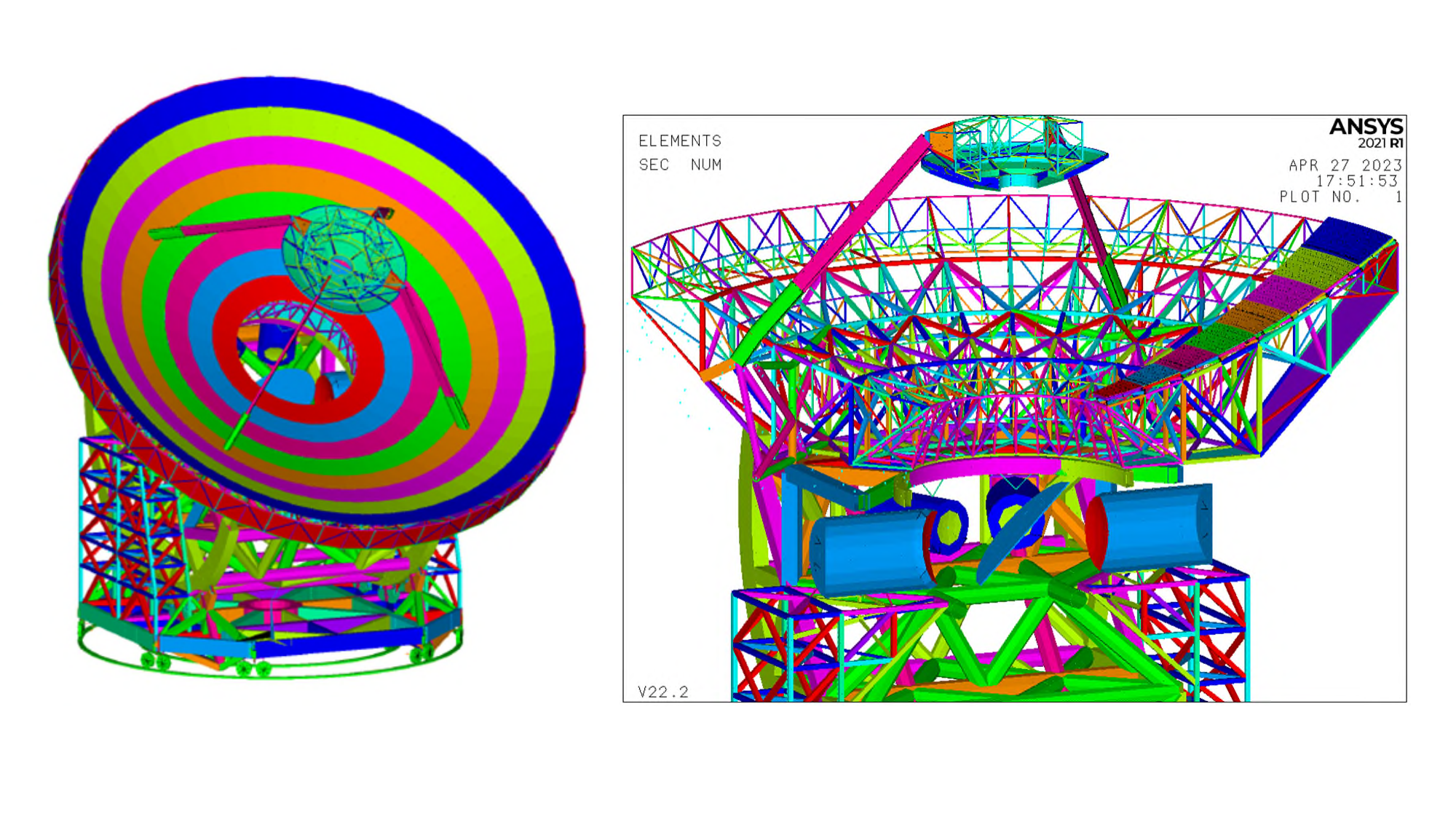}
    \caption{Finite element model of the full telescope. Left: full model view. Right: cut view along the elevation axis}
    \label{fig:fe model}
\end{figure}

\begin{figure}[htb]
    \centering
    \includegraphics[trim={0.0cm 2.0cm 0.0cm 2.0cm},clip, width=1\linewidth]{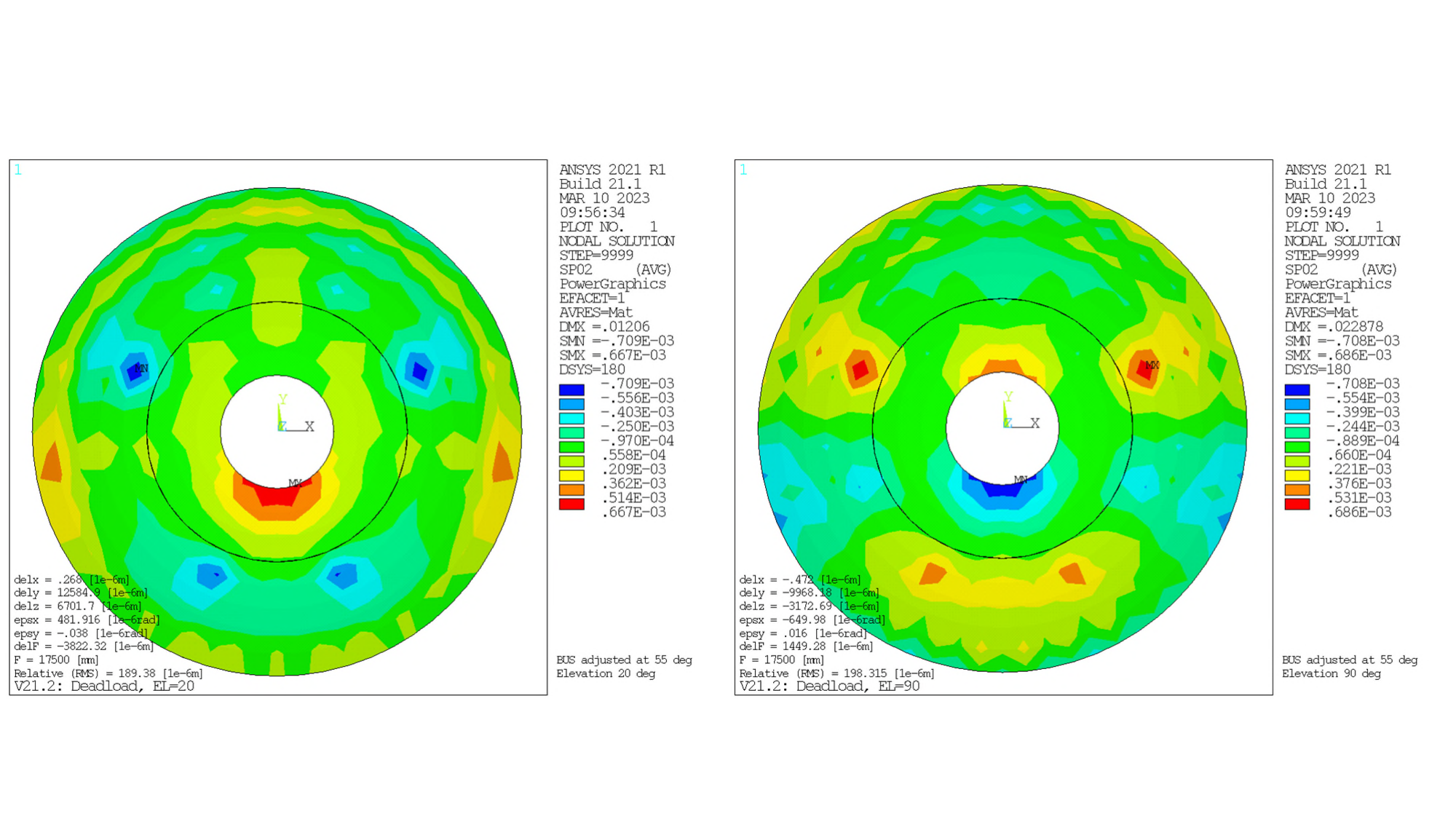}
    \caption{M1 surface deformation under gravity at elevation positions of 20$^\circ$ and 90$^\circ$, demonstrating the compliance with the requirement of $< 200~\mu$m rms (before active corrections). }
    \label{fig:M1 surface plot}
\end{figure}

The main reflector design follows a homology principle that has been applied on other comparable telescopes with steel reflectors where the backup structure has a iso-static four point interface with the elevation wheel structure. This results in a advantageous deformation behavior of the surface under changing gravitation vectors. During the project a full finite element model was developed using  Ansys$^\textsc{TM}$ 2021 R1 software suite.\footnote{\url{https://www.ansys.com/}} to demonstrate the compliance of the structure with values down to 200~$\mu$m rms HWFE without an active surface (see Figures~\ref{fig:fe model} \& \ref{fig:M1 surface plot}). 

In order to achieve a lightweight and stiff design with low thermal expansion for the 12m secondary mirror M2, a high-modulus carbon fiber reinforced plastics (HM-CFRP) is used for the backup structure, being inspired by the ALMA antenna main reflector design. M2 is mounted on an active hexapod arrangement to correct any misalignments in the optical path. 

The tertiary mirror M3 has an elliptical shape with a short axis of 6~m and a long axis of 8.6 m. Again a CFRP backup structure is foreseen that is iso-statically mounted on a rotation mechanism. For both M2 and M3 it is planned to use the same tile concept as for M1 (electroformed nickel or machined aluminum tiles, see \ref{sec:active main reflector}) that are attached to the base plate by four adjusters transferring axial loads and one adjuster transferring lateral loads.

\begin{figure}[tbh]
    \centering
    \includegraphics[trim={0.0cm 4.5cm 0.0cm 3.5cm},clip,width=1\linewidth]{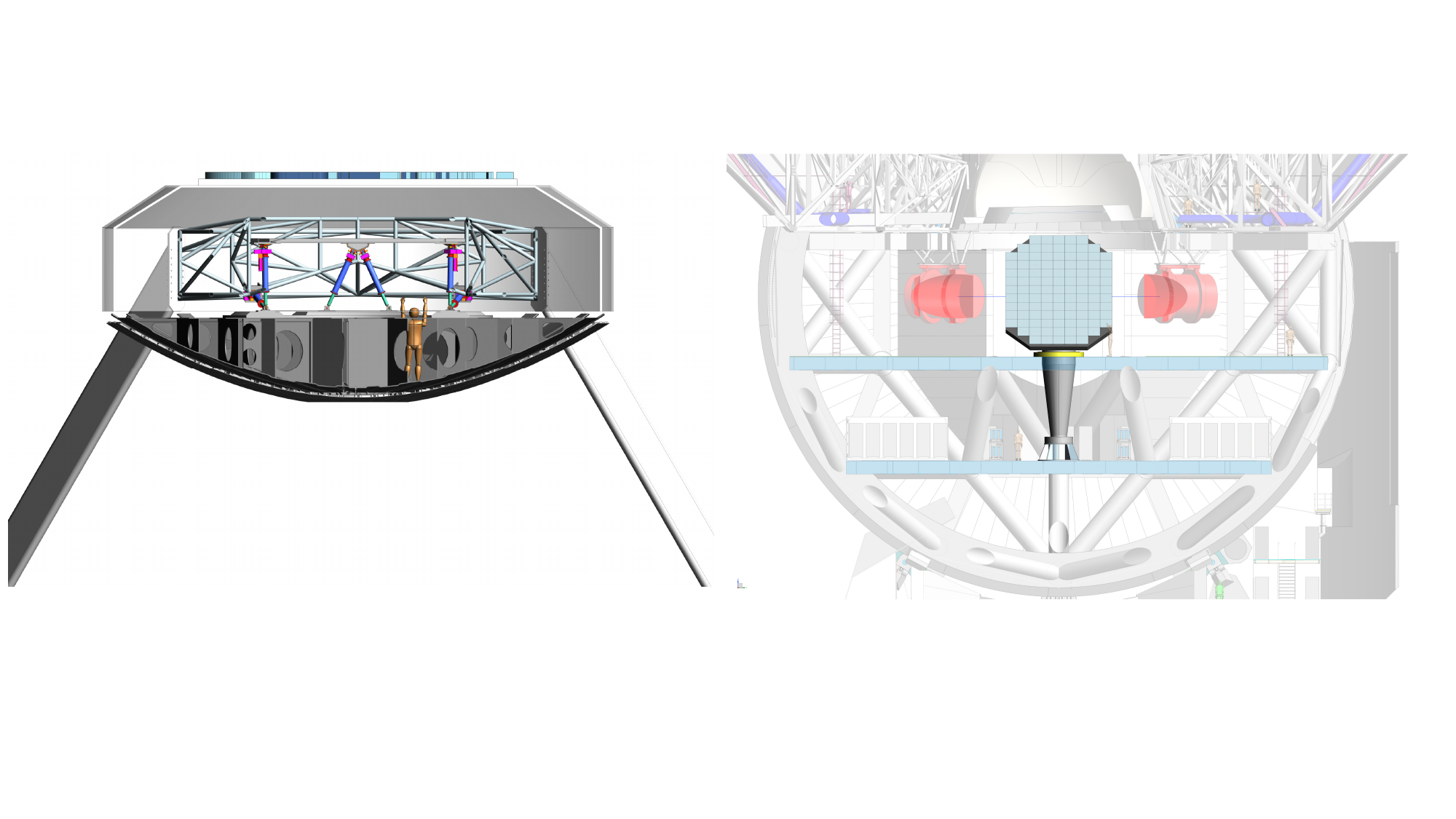}
    \caption{Concepts for the secondary reflector M2 and tertiary reflector M3.}
    \label{fig:M2 and M3}
\end{figure}

\subsection{Main Reflector Surface}
\label{sec:active main reflector}
Key to meeting the high HWFE requirements for local deformations of the main reflector surface is the layout of the reflecting individual surface elements, in this case the panel-segments. The design concept for the panel-segments was influenced by the segmented reflectors of the LMT and the Keck telescopes, having iso-static features as used for optical mirrors. The use of panel segments offers several advantages. On the one hand, the assembly and adjustment effort is considerably reduced. On the other hand, the segments can be mechanically decoupled from the underlying BUS structure by means of actuators at the connection point which compensate the major sag of the BUS under dead load. 
We note that Puddu et al., submitted to these proceedings, analyzes the impact of the panel-segment gaps on the expected optical performance of AtLAST, finding the impact is acceptable and subdominant to the overall HWFE.
While the basic principle of segmentation was adopted from the LMT, some new developments are applied. The size of one panel segment was adjusted to 2.5 meters to facilitate handling of the units. This results in eight concentric rings with a total of 416 segments,\footnote{For comparison, we note that the European Extremely Large Telescope will comprise 798 segments.} with identical shapes in each ring. 

It is necessary to further split the surface of one panel segment into smaller panel tiles to allow the unavoidable thermal strain under thermal load to be accepted without compromising the required accuracy. 
When splitting one panel-segment into several panel tiles, a compromise must be made between the size dependent manufacturing accuracy of the reflector tiles and the assembly and adjustment effort depending on the number of tiles. For reasons of accuracy, milled aluminum or electroformed nickel tiles, that have been used for the ALMA antennas, can be used as a manufacturing driven design approach. The trade-off resulted in edge lengths between 0.6m  and 0.8m, being grouped into panel-segments carrying between 12 and 16 tiles.
The panel tiles are mounted via adjuster elements (rods) on a baseplate (CFRP honeycomb sandwich or a monocoque type CFRP grid). Four rods per tile carry the loads normal to the surface and one rod the lateral loads, providing directional decoupling of the load reaction.
The baseplate rests on a steel subframe that is torsionally soft to avoid statically overdetermined connection of the subframe in surface normal direction to the BUS.
Finally, linear actuators connect the subframe to the main reflector backup structure.

\begin{figure}[tbh]
    \centering
    \includegraphics[trim={4.0cm 0.0cm 4.0cm 0.0cm},clip, width=0.8\linewidth]{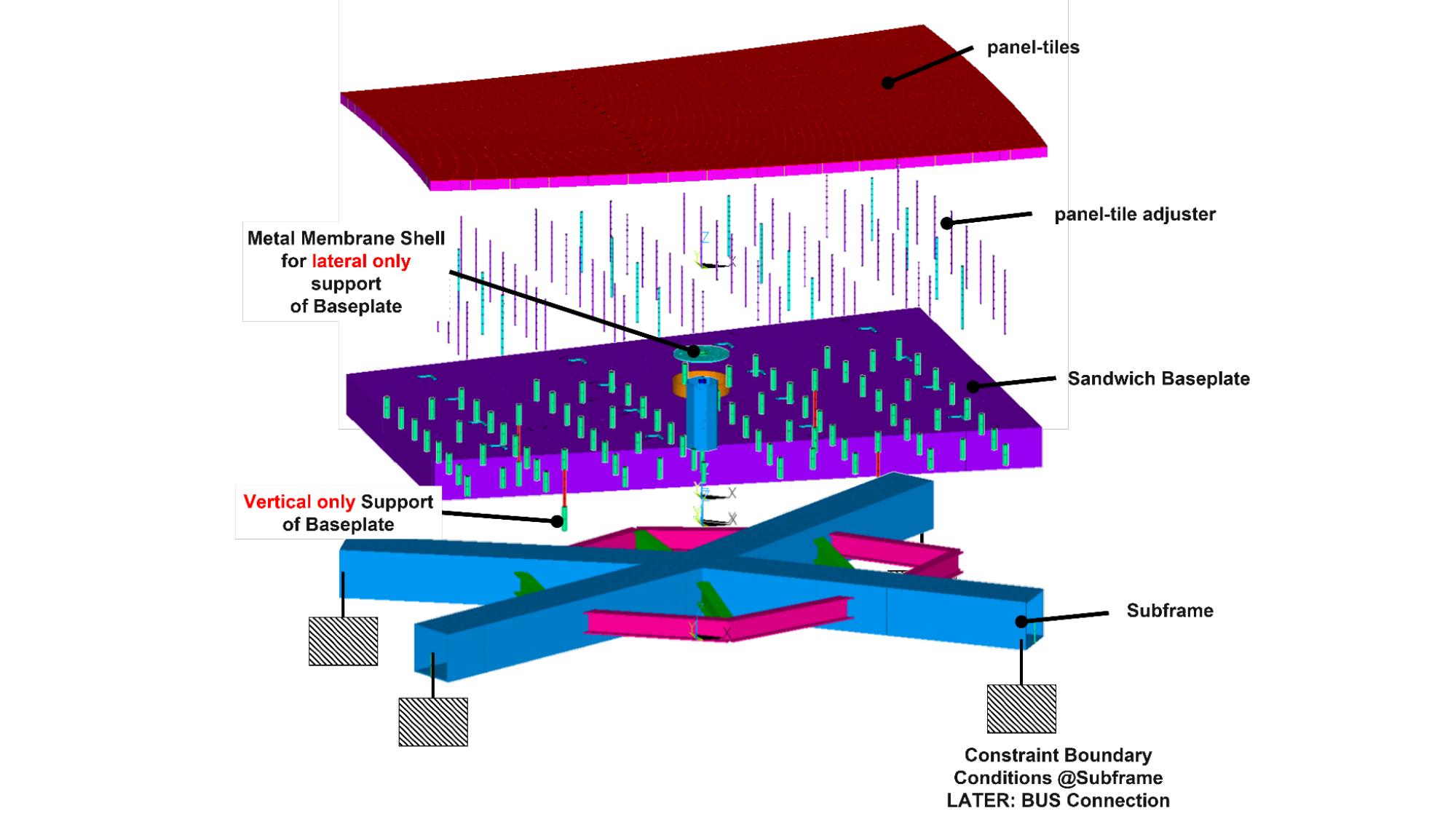}
    \caption{Simplified panel segment subsystem model}
    \label{fig:panel segment}
\end{figure}

\begin{figure}[tbh]
    \centering
    \includegraphics[trim={4.0cm 0.0cm 4.0cm 0.0cm},clip, width=0.75\linewidth]{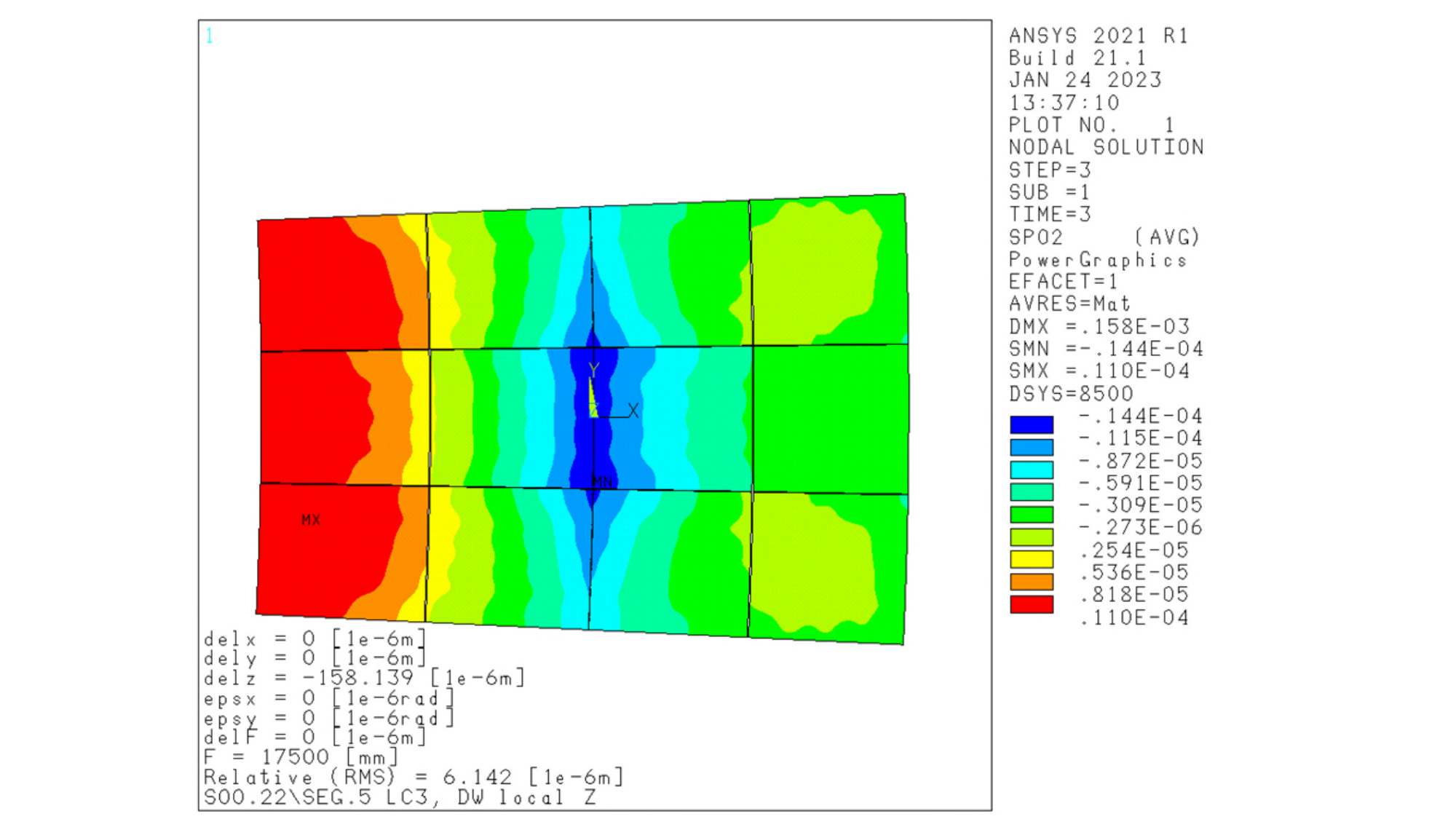}
    \caption{Example panel-segment surface plot. The maximum deformation is 6.1~$\mu$m rms at EL = 90$^\circ$}
    \label{fig:panel_deform}
\end{figure}

To verify the panel segment contribution to the HWFE budget, simplified finite element subsystem models were created (see Fig.~\ref{fig:panel segment}). The deformation behavior was investigated for changing gravity vectors. Fig.~\ref{fig:panel_deform} shows an exemplary surface plot verifying the compliance with the requirement of $< 8~\mu$m rms from the HWFE budget table.

\subsection{Active Optics (AcO)}
Considering the HWFE budget in Fig.~\ref{fig:HWFE Budget}, the largest, yet also the most repeatable, errors can be expected from the gravitational deformation of the primary reflector backup structure (BUS) due to changing elevation angle and thus a changing angle of attack of the gravity vector. The passive HWFE rms due to the BUS gravitational deformations is targeted to be 200~$\mu m$ for AtLAST (and can be reached as demonstrated by finite element analysis). Since this effect is very repeatable, a passive system using lookup tables can serve as the first step in yielding a significant accuracy improvement depending on the accuracy of the measurement method used for the calibration (photogrammetry, holography on celestial sources). This is a proven method in existing telescopes with active surfaces such as the Sardinia Radio Telescope (SRT) and the Large Millimeter Telescope Alfonso Serrano (LMT).
But looking again in Fig.~\ref{fig:HWFE Budget}, there are contributions from slowly changing, transient temperature distributions and wind (quasi-static, steady-state, excitation frequency bandwidth $\leq 0.1$~Hz) that can be only observed and compensated when using a dedicated sensor system in a closed feedback loop with the M1 surface actuators. The contribution of those slow transient environmental loads is still in a magnitude range that would otherwise prevent the M1 surface from reaching the required accuracy.

As a consequence, for a closed loop feedback-control of the active main reflector surface, a potent sensor system is crucial to detect the actual position of the panel segments. 
The measuring accuracy of the sensor system needs to be an order of magnitude better than the positioning requirement, i.e.\ $< 1~\mu$m. A challenging figure for a 50~m reflector.
A technology assessment was conducted to identify suitable solutions. Proven photogrammetric or laser tracker methods cannot reach the required accuracy.

\begin{figure}[htb]
    \centering
    \includegraphics[trim={1.0cm 0.0cm 0.5cm 0.0cm},clip, width=0.99\linewidth]{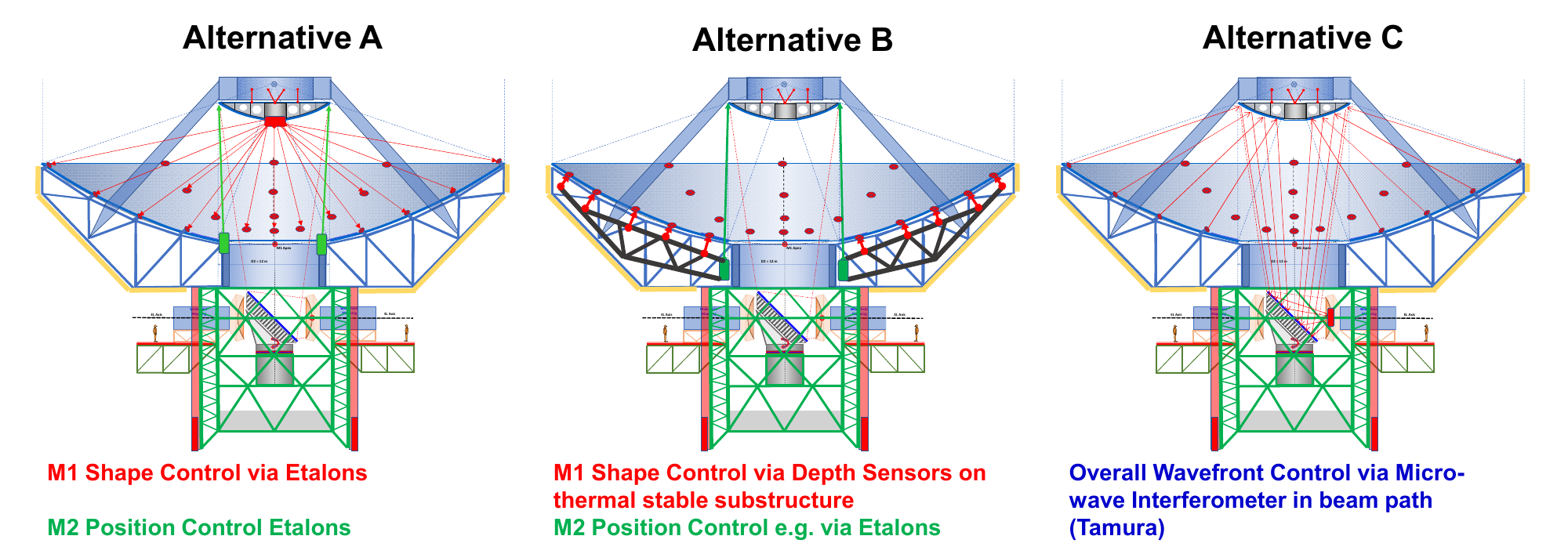}
    \caption{Alternatives for a metrology system}
    \label{fig:metrology}
\end{figure}

The only optical system with adequate technology readiness level turned out to be the Etalon Absolute Multiline Technology$^\textsc{TM}$. It is already used  in optical telescopes such as ESO’s Very Large Telescope (VLT), the Large Binocular Telescope (LBT) and the future Giant Magellan Telescope (GMT; Rakich et al.\ 2016)\cite{Rakich2016}. The Sardinia Radio Telescope (SRT) is currently commissioning a system for the relative alignment of the M2 with respect to M1, as described in Attoli et al.\ 2023.\cite{Attoli2023} The Etalon Absolute Multiline$^\textsc{TM}$ system achieves a nominal accuracy/uncertainty of $0.5~\mu \rm m \cdot m^{-1}$ in air according to the datasheet. A network of those sensors can detect the relative position between the optical elements M1, M2 and M3 as well as the absolute position of one of the elements, see Fig.~\ref{fig:metrology}). 
The exposure of the sensors needs to be considered in terms of reliability and calibration measures.
A planned system to measure the M1 surface deformations of the Large Millimeter Telescope Alfonso Serrano (LMT) was described in Schloerb et al.\ 2022 \cite{Schloerb2022}.

An alternative solution is the use of distance (depths) sensors on a thermally stable CFRP structure  within the M1 BUS (see Fig.~\ref{fig:metrology} B). The position of M2 can be realized again with Etalon sensors. The upside of such a system is the controlled environment around the sensor systems and thus a high expected reliability of the system.

A entirely different approach is wavefront control via a microwave interferometer in the beam path as developed by Tamura et al.\ 2020\cite{Tamura2020} and Nakano et al.\ 2022\cite{Nakano2022} for the Large Submillimeter Telescope (LST) conceptual design, and tested on the 45-meter Nobeyama telescope (see Fig.~\ref{fig:metrology}C).
While it measures the whole beam path from M1 via M2 and M3 to the science instruments, it becomes difficult to separate between the contributions from positional errors in the reflectors.

Overall, it is likely that AtLAST would benefit by further investigations of more than one of the alternatives presented in this section. This is an area of active development to be explored in the next phase of the AtLAST project.

\section{Pointing accuracy}

\subsection{Blind Pointing Accuracy}

\begin{figure}[htb]
    \centering
    \includegraphics[trim={0.5cm 0.2cm 0.5cm 0.2cm}, clip, width=0.99\linewidth]{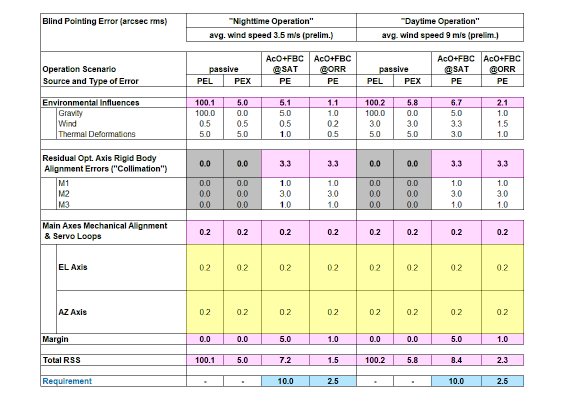}
    \caption{Blind pointing error budget}
    \label{fig:blind pointing}
\end{figure}

The blind pointing error is the residual deviation from the desired pointing direction of the telescope after an arbitrary slewing operation (moving the telescope to a new target/object between measurements) within the celestial hemisphere. 
Fig.~\ref{fig:blind pointing} shows the blind pointing error budget. The overall blind pointing error shall not exceed 2.5\arcsec\ at low wind nighttime operations.
Contributions to the blind pointing error by the telescope structure can be classified as:
\begin{itemize}
\item systematic, predictable deformations and misalignment (such as changing gravity vector with respect to the elevation direction)
\item systematic, observable (by sensors), but non-predictable, slow (low frequency, $\leq0.1$~Hz) transient deformations due to temperature distributions, steady state-wind.
\item purely random or non-observable, non-predictable disturbances (mostly higher frequency, $>0.1$~Hz) due to wind, internal disturbances (motor cogging dynamic friction effects, vibrations of auxiliary machinery).
\end{itemize}

Systematic, predictable errors occur due to gravitational deformations of the structure and repeatable misalignment from manufacturing and assembly tolerances (e.g.\ deviation of the azimuth axis from the vertical direction, non-perpendicularity of azimuth- and elevation axis etc.) that purely depend on the pointing coordinates (altitude/elevation and azimuth). These deformations are predictable and (luckily) represent the largest contributions to the overall blind pointing error. Compensation approaches can involve a classical, fitted pointing model from pointing samples or a machine learning approach considering more input data than just the coordinates and errors of the pointing samples, as discussed in Thoms et al.\ submitted in these proceedings.
The blind pointing error budget's column labeled ``passive'' in Fig.~\ref{fig:blind pointing} shows a state where no compensation methods are applied.
Therefore, the magnitude of the uncorrected pointing error in elevation direction due to gravitational deformations shows a very high values of 100~arcsec.

Systematic, observable, but unpredictable, slow (low frequency, $\leq 0.1$~Hz) transient deformations
are caused by external environmental loads such as wind and temperature distributions inside the structural materials. 
They can be continuously observed by sensor systems and compensated by actuators –- what we call the metrology system here. 
The subset of the metrology functions/sensors aiming at the compensation of all deformations with respect to pointing (except components in the optical path) is defined as the aforementioned ``flexible body compensation'' (FBC).
The FBC exclusively controls the telescopes pointing direction (controlled quantity). 
Its actuators are solely the telescope main axes.
The benefit of the active systems over the static pointing error model in Fig.~\ref{fig:blind pointing} in column ``AcO+FBC @SAT" is mainly the reduction thermal deformations (which is not too high during nighttime condition) and average wind (which is also low in the case of low wind speed). 
The FBC will be coupled with the AcO since corrections in both systems interact with pointing. 

Residual blind pointing errors after the correction by static pointing error models and FBC under final operational conditions are caused by purely random, not observable nor predictable, fast disturbances (frequency, $>0.1$~Hz). They do originate from the high frequency fraction of wind gust, imperfections of drive torque and transmissions and random mechanical errors (bearings). They represent the accuracy limit which must satisfy the science requirements.

\subsection{Steady State (Constant) Wind-Induced Blind Pointing Error}
\begin{figure}[htb]
    \centering
    \includegraphics[width=0.8\linewidth]{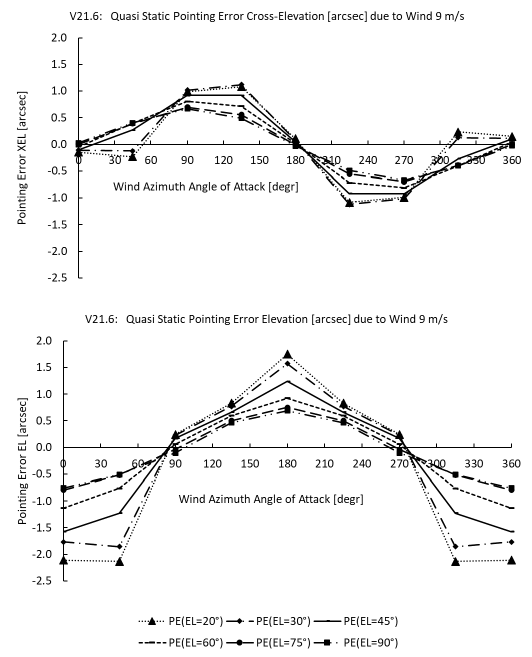}
    \caption{
    Upper: Pointing Error in Cross-Elevation Direction (perpendicular to momentary elevation direction).
    Lower: Pointing error in EL due to continuous, quasi-static wind-load 9 m/s (one line per elevation position).}
    \label{fig:Pointing curves wind}
\end{figure}

The steady state wind enduced pointing error was investigated with the full FE model for different telescope positions and wind attack angles.
The curves in the upper panel of Fig.~\ref{fig:Pointing curves wind} show the wind-induced blind pointing error in the cross-elevation direction only.\footnote{For reference, cross-elevation is defined as the direction orthogonal to the elevation direction when considering the image tangent plane.} As expected by intuition, the pointing error in cross-elevation direction has a maximum when the wind blows from the side (wind azimuth angle of attack 90$^\circ$ or 270$^\circ$) while it has a minimum when the wind attacks the dish from the front.  

The curves plotted in the lower panel of Fig.~\ref{fig:Pointing curves wind} show the steady state (constant) wind pointing error (static load case) over the wind azimuth angle of attack for different elevation angles. The pointing error shown is only the elevation component of the pointing error. As expected by intuition the pointing error in elevation direction has a maximum when the wind blows directly on the reflector front side (wind azimuth angle of attack 0$^\circ$) or on the rear side (wind azimuth angle of attack 180$^\circ$). The minimum wind induced blind pointing error in elevation can be expected when the wind blows from the side relative to the pointing direction.

The values shown for the wind elevation and cross elevation pointing error represent a very good pointing performance for the wind speed and size of the structure given. 

\subsection{Pointing Stability}
\begin{figure}[htb]
    \centering
    \includegraphics[trim={0.6cm 0.2cm 0.6cm 0.2cm}, clip, width=0.8\linewidth]{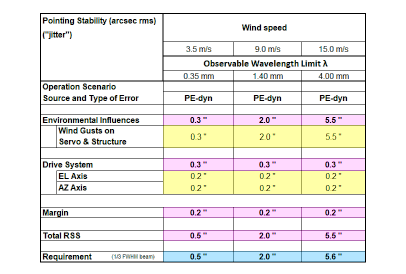}
    \caption{Pointing stability budget}
    \label{fig:pointing stability}
\end{figure}

During the observation of a celestial object when the science instruments collect or integrate source signals, the pointing direction on sky must be maintained over a defined time interval (the longest integration time of science instruments).
The tracking accuracy or tracking error is the rms deviation over this interval from the initial pointing direction on source. The tracking accuracy is mainly impacted by small random dynamic errors (wind gusts, friction jitter, motor cogging) and drift over time by the same root causes as the residual errors after corrections of the blind pointing error.
The driving factor for AtLAST's required tracking accuracy is the desired signal to noise ratio of the instruments at the best, diffraction limited resolution of the telescope optics. While the airy disk at a wavelength of $350~\mu$m is roughly 1.8\arcsec\ (see, e.g., Puddu et al.\ submitted in this volume), the instruments require a value of 0.5\arcsec\ over an integration interval to acquire enough signal.
The requirement was then broken down to the engineering budgets that can be seen in Fig.~\ref{fig:pointing stability}.

The pointing stability budget includes mainly random disturbances acting on the structure and thus the control loop and a tiny fraction of drift accumulating from remaining systematic errors after application of all blind pointing corrections. 
This explains why the contribution values are small compared to the blind pointing error budget.
The major contributor however will be wind gusts. While observations at short wavelengths down to 350~$\mu$m at low wind speeds up to 3.5~m~s$^{-1}$ define the critical tracking stability value of 0.5\arcsec\ rms, the tracking jitter will conservatively scale with the square of the wind speed. Since the instruments demand for tracking stability is a function of the observed wavelength, the maximum observable wavelength will be limited by the average wind speed.

\section{CONCLUSION}
The AtLAST telescope will enable us to map and image the submillimeter sky at unprecedented speeds, and at a resolution well-matched to premier current and next generation wide field low frequency radio and optical facilities. It will be the first subarcminute resolution (sub-)mm telescope to provide a field of view in the range of 1-2$^\circ$, at the remarkable level of sensitivity provided by the 50 meter aperture. Both Nasmyth and Cassegrain instrument mounts are provided in order to accommodate the needs of the instrument builders.

The performance metrics and engineering approaches which follow from the breakdown of the top-level requirements resemble those of optical telescopes more than those of classical millimeter and submm-wave telescopes.
However, in contrast to the majority of optical telescopes, the structure of the AtLAST telescope is too large to be protected from the environment by a dome. Consequently, for the first time when considering a new, large (sub-)mm telescope, a closed feedback-loop metrology system for the optical surfaces is deemed necessary from the beginning in order to achieve the needed surface-, alignment- and pointing accuracy (not considering telescopes using calibrated look-up tables such as the active surfaces of the GBT, SRT and LMT). 

This paper focused on the breakdown of the top-level requirements to engineering budgets in the first phase of the design study. 
Here, the general approach for the telescope design development largely followed proven concepts from existing telescopes.
Doing so, the technical risks for the majority of the subsystems could be limited (e.g.\ the wheel on track concept for the azimuth axis mechanism and the CFRP backup structure for the secondary mirror).
Our proof of concept has been elaborated for many of the subsystems contributing to the error budgets e.g.\ via finite element model simulations.
For other subsystems the technology readiness level still needs to be increased in the next project phases (e.g.\ the metrology concept with sensors). For more detailed description of the subsystems, see Mroczkowski et al.\ 2024 \cite{Mroczkowski2024}.

Overall, the AtLAST design study has delivered a robust conceptual telescope design fulfilling the top level requirements that can be realized after verifying the technologies in a next project phase using breadboard and field tests on existing facilities. Following this, we confirm that -- if fully funded --AtLAST could be begin construction later this decade.

\acknowledgments 

This project has received funding from the European Union’s Horizon 2020 research and innovation program under grant agreement No.\ 951815 (AtLAST).\footnote{See
\href{https://cordis.europa.eu/project/id/951815}{https://cordis.europa.eu/project/id/951815}.}   
The consortium consists of the University of Oslo, the European Southern Observatory, OHB Digital Connect GmbH (formerly MT Mechatronics), the United Kingdom Astronomy Technology Centre (UKATC), and the University of Hertfordshire.

\bibliography{report} 
\bibliographystyle{spiebib} 

\end{document}